\documentstyle[graphicx]{mn}

\newif\ifAMStwofonts
\AMStwofontstrue

\def\pg{{PG1211+143}}

\def\msun{{\rm M_{\odot}}}

\def\mo{{\dot M_{\rm out}}}

\def\xmm{{\it XMM-Newton}}

\def\et{{et al.\ }}

\newcommand{\ls}{\mathrel{\hbox{\rlap{\hbox{\lower4pt\hbox{$\sim$}}}\hbox{$<$}}}}
\newcommand{\gs}{\mathrel{\hbox{\rlap{\hbox{\lower4pt\hbox{$\sim$}}}\hbox{$>$}}}}

\def\arcs{{\hbox{$^{\prime\prime}$}}}

\def\Msun{\hbox{$\rm ~M_{\odot}$}}

\def\H0{{\rm ~km~s^{-1}~Mpc^{-1}}}

\def\msun{M_{\rm \odot}}

\def\et{{et al.}}

\title[High velocity outflow of \pg]
        {Confirming the high velocity outflow in \pg}
\author[K.A.Pounds \et]
        {K.A.Pounds, K.L.Page\\
Department of Physics and Astronomy, University of Leicester,
Leicester, LE1 7RH, UK\\}

\date{Accepted ; Submitted }
\pagerange{\pageref{firstpage}--\pageref{lastpage}}
\pubyear{2005}
\begin{document}
\maketitle
\label{firstpage}

\begin{abstract}
An \xmm\ observation of the bright QSO \pg\ was previously reported to show evidence for a massive, energetic outflow, with an outflow velocity of
v$\sim$0.1c based on the identification of blue-shifted absorption lines detected in both EPIC and RGS spectra. Subsequently, an order-of-magnitude
lower velocity has been claimed from an ion-by-ion model fit to the RGS data. We show here, in a re-analysis of the higher signal-to-noise EPIC
data, that the high velocity is confirmed, with the resolution of additional absorption lines yielding a revised outflow velocity in the range $\sim$0.13-0.15c. 
Confirmation of a massive and energetic outflow in a non-BAL AGN has important implications for metal enrichment of the IGM and
for the feedback mechanism implied by the correlation of black hole and galactic bulge masses. We note the near-Eddington luminosity of \pg\ may be
the critical factor in driving such an energetic outflow, a situation likely to be common in AGN at higher redshift.    

\end{abstract}


\section{Introduction}

An analysis of the  EPIC and RGS spectra from an \xmm\ observation of the bright narrow emission line QSO \pg\ in 2001 provided evidence for a highly
ionised outflow with a velocity of $\sim$0.1c (Pounds \et\ 2003; hereafter P03), though a lower velocity has recently been claimed from a separate
analysis, principally based on the low signal-to-noise RGS data (Kaspi and Behar 2006). Confirmation of the high velocity outflow  is important since
the mechanical energy in the flow, if not highly collimated, is a significant fraction of the bolometric luminosity of \pg\ and could be typical of
AGN accreting near the Eddington rate (King and Pounds 2003), while also providing an example of the feedback required by the linked growth of SMBH
in AGN with their host galaxy (King 2005). Subsequently, the same \xmm\ observation of \pg\ has been used by Gierlinski and Done (2004) to suggest
how strong absorption of the intrinsic X-ray continuum in a `velocity-smeared', high column, of moderately  ionised gas can provide a physically
preferred explanation (to Comptonisation) for the strong soft excess widely seen in type 1 AGN (Wilkes and Elvis 1987, Turner and Pounds 1989). A
similar study by Chevallier \et\ (2006), which also considered an ionised reflection origin of the soft excess, concluded that absorption was the
more likely cause of a strong soft excess (as in \pg).

In the present paper we re-examine the \xmm\ EPIC data of \pg\ which formed the strongest evidence for the high velocity claimed in P03. We use the
latest calibration files and - in particular - take advantage of the higher spectral energy resolution of the MOS cameras demonstrated in recent
studies of the Type 2 Seyferts Mkn3 (Pounds and Page 2005) and NGC1068 (Pounds and Vaughan 2006).  

We assume a redshift for \pg\ of $z=0.0809$ (Marziani \et\ 1996). 

\section{Observation and data reduction}

\pg\ was observed by \xmm\ on 2001 June 15 for $\sim$53 ks. In this paper we concentrate on data from the EPIC pn (Str\"{u}der \et 2001) and MOS (Turner
\et 2001) cameras, which were deployed in large and small window mode, respectively, both with the medium filter. All X-ray data were first screened with the XMM SAS v6.5 software and events corresponding to patterns 0-4 (single and double pixel
events) selected for the pn camera and patterns 0-12 for the MOS1 and MOS2 cameras. We extracted source counts within a  circular region of 45\arcs\
radius defined around the centroid position of \pg, with the background being taken from a similar region, offset from but close to the source. After
removal of data during periods of high background the effective pn exposure was $\sim$49.5 ks while the MOS cameras were combined to give a
single-camera-equivalent exposure of $\sim$107 ks. Individual spectra were binned to a minimum of 20 counts per bin, to facilitate use of the
$\chi^2$ minimalisation technique in spectral fitting. 

Spectral fitting was based on the Xspec package (Arnaud 1996), version 11.3. All spectral fits include absorption due to the line-of-sight Galactic
column of $N_{H}=2.85\times10^{20}\rm{cm}^{-2}$  (Murphy \et\ 1996) and errors are quoted at  the 90\% confidence level ($\Delta \chi^{2}=2.7$ for
one interesting parameter).

\section{Absorption lines in the EPIC data}   

Figure 1 reproduces the ratio plot of pn data to a simple power law fit (photon index of $\Gamma$$\sim$1.78) over the 1--10 keV band, as modelled in
P03. Marked on the figure are 3 statistically  significant `narrow' absorption lines, with their identification as proposed in P03, where they
formed a key part of the case for a high velocity outflow. Denoted by an asterix in figure 1 is a further `absorption line' at $\sim$1.8 keV, which was
not included in the P03 analysis due to prevailing uncertainties in the detector calibration near the neutral Si edge. Repeating the analysis in P03
by successively adding gaussian lines to the
power law model in Xspec the observed absorption lines can be quantified. Fitting first the strongest line at $\sim$7 keV improves the fit by $\Delta$ 
$\chi^{2}$ of 69 for 3 fewer degrees of freedom,
with a line energy of 7.07$\pm$0.03 keV (observer frame), width $\sigma$=168$\pm$46 eV and flux -5.7$\pm$0.9 $\times 10^{-6}$ ph cm$^{-2}$ s$^{-1}$, corresponding to
an equivalent width against the power law continuum of $\sim$210$\pm$35 eV.  Fitting additional gaussians to the weaker spectral features at $\sim$2.7
and $\sim$1.5 keV, with line width fixed at $\sigma$=50 eV, gives further significant improvements to the fit, with $\Delta$$\chi^{2}$ of 17 and 16,
respectively, each for 2 additional degrees of freedom. The best fit line energies, again in the observer frame, are 2.7$\pm$0.03 and 1.47$\pm$0.02 keV, with respective fluxes of 
-4.5$\times 10^{-6}$ ph cm$^{-2}$ s$^{-1}$ and -5.7$\times 10^{-6}$ ph cm$^{-2}$ s$^{-1}$ (EWs of $\sim$30eV and $\sim$14eV, accurate to a factor $\sim$2). Identifying the 3
absorption
lines with the resonance Ly$\alpha$ transitions of FeXXVI, SXVI and MgXII then yields the mean outflow velocity of $\sim$0.09$\pm$0.01c reported in
P03.

\begin{figure}                                                          
\centering                                                              
\includegraphics[width=6cm, angle=270]{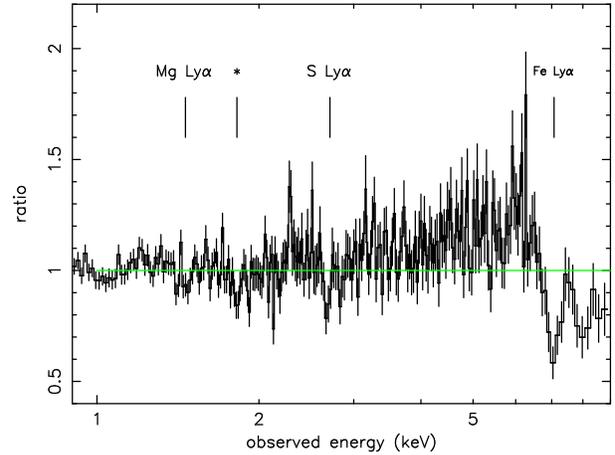}                                          
\caption                                                                
{EPIC pn camera data from the observation of \pg\ in 2001 compared with a simple power law fit over the energy band 1-10 keV. Three narrow 
spectral features in the 2001
data and their proposed identification in P03 with absorption lines of highly ionised Mg, S and Fe are indicated. A further significant `absorption 
line', marked by an asterix, lies close to the neutral Si absorption edge and was therefore ignored in P03}      
\end{figure}

\begin{figure}                                                          
\centering                                                              
\includegraphics[width=6cm, angle=270]{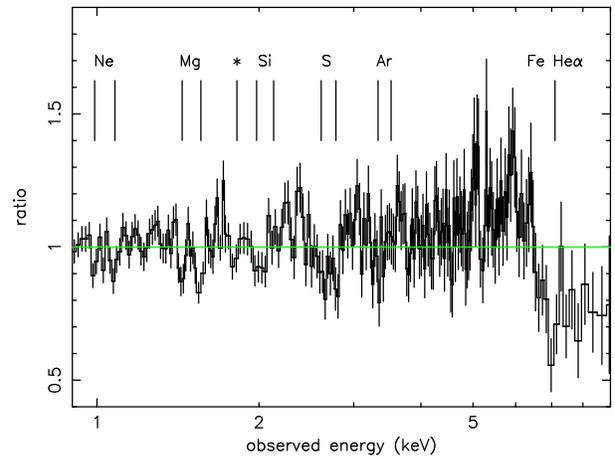}                                          
\caption                                                                
{EPIC MOS camera data from the observation of \pg\ in 2001 compared with a simple power law fit over the energy band 1-10 keV. Narrow spectral features 
and their proposed identification with K-shell absorption lines of Ne, Mg, Si, S, (Ar?) and Fe are indicated}      
\end{figure}

\subsection{Examining the higher resolution MOS spectrum}
As noted in P03, the outflow velocity of $\sim$0.09c is conservative in that identifying the $\sim$7 keV feature with Fe XXV He$\alpha$,
rather than  FeXXVI Ly$\alpha$, would require a larger energy shift and correspondingly higher velocity. Encouraged by the impressive low energy spectra from
the (higher resolution)  MOS cameras, demonstrated in two recent studies of the Seyfert 2 galaxies Mkn3 (Pounds and
Page 2005) and NGC1068 (Pounds and Vaughan 2006) we have re-examined the spectral structure in the \xmm\ observation of \pg, concentrating on the
MOS data. Our aim was to clarify the spectral structure at medium and low energies in the hope of removing the ambiguity in identifying the $\sim$7 keV
absorption line, and hence  improving confidence in the deduced outflow velocity.

\begin{figure}                                                          
\centering                                                              
\includegraphics[width=4.7cm, angle=270]{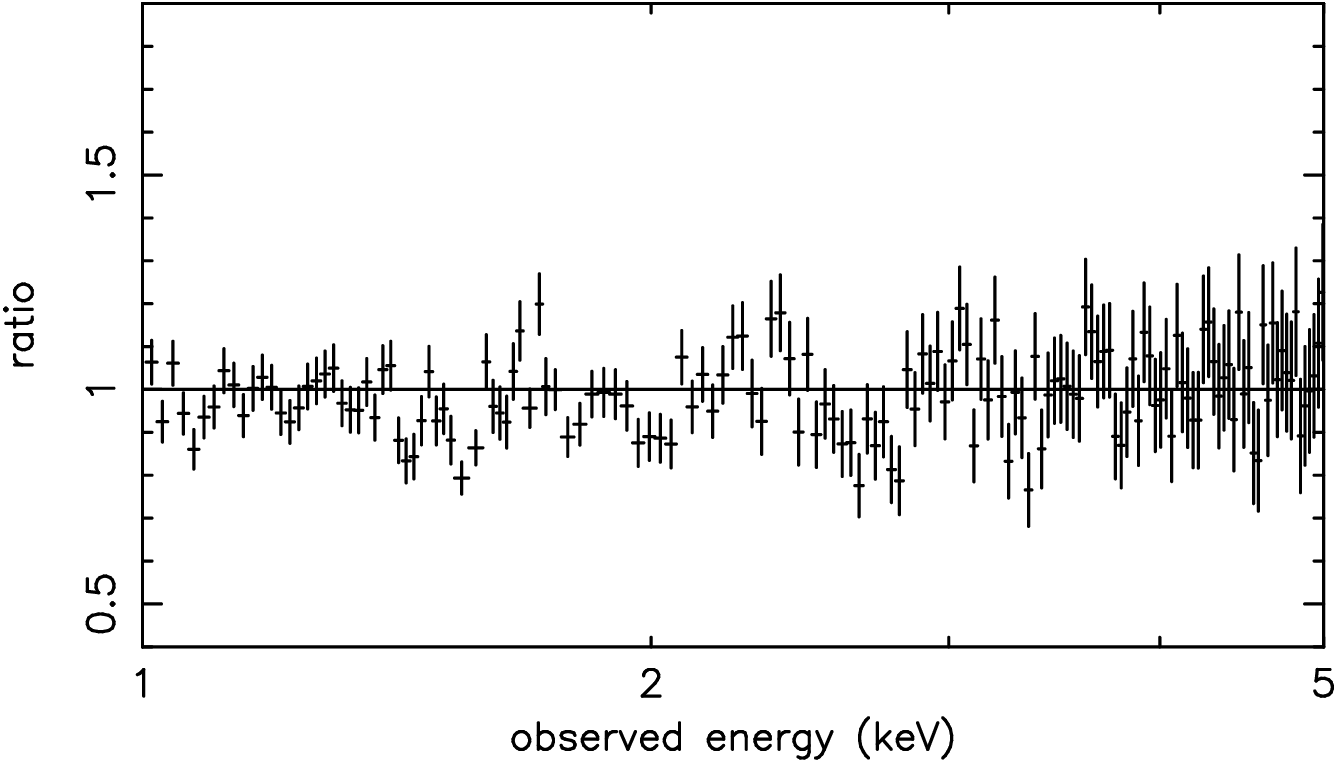}                                          
\centering                                                              
\includegraphics[width=4.7cm, angle=270]{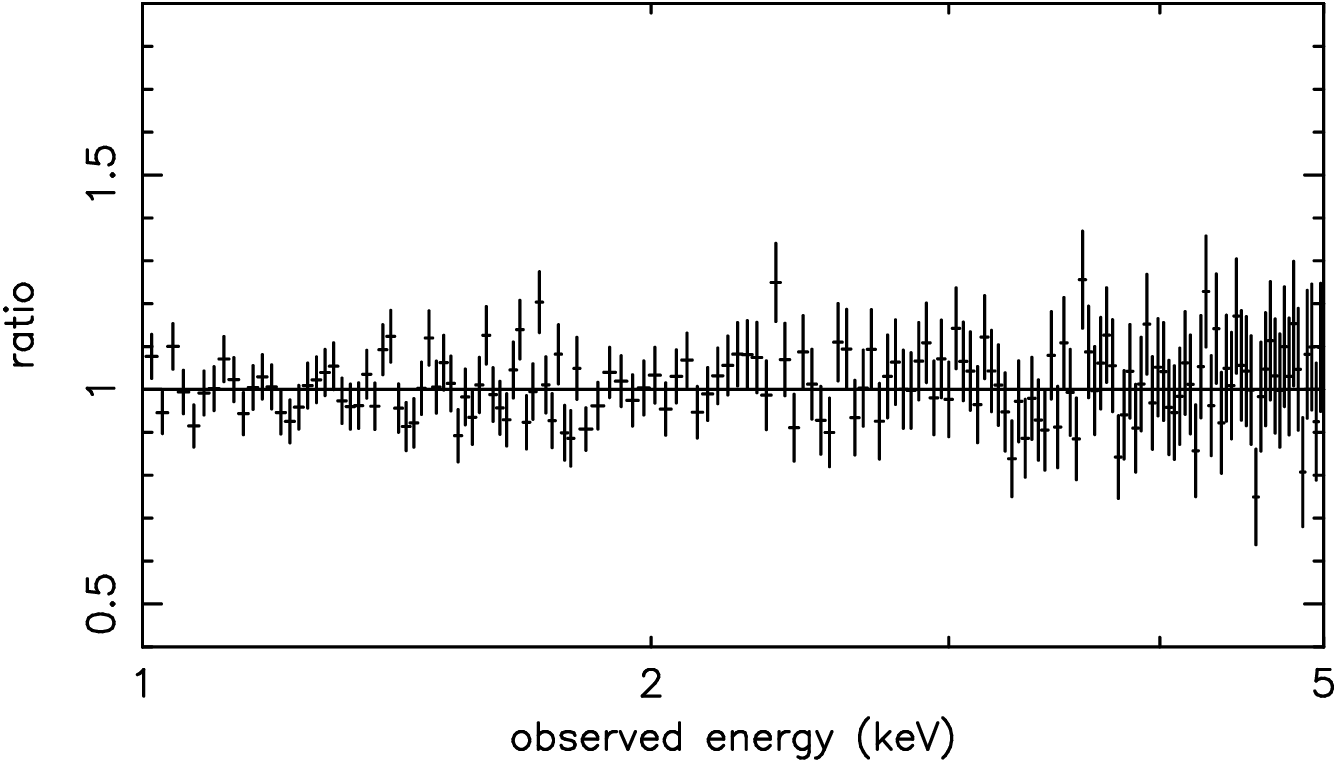}                                          
\caption                                                                
{(Upper panel) Ratio of MOS camera data to a simple power law fit over the energy band 1-5 keV. Several narrow spectral features 
contibute to a statistically poor fit. (Lower panel) Ratio plot after inclusion of 6 absorption lines with the parameters listed in Table 1
yielding a reduction in $\chi^{2}$ of 81 for 12 fewer degrees of freedom}      
\end{figure}

\begin{table*}
\centering
\caption{The 6 strongest absorption lines identified in the MOS spectrum of \pg. Line energies are in keV and give the observed and laboratory values, and their 
ratio. The fitted line
fluxes are units of $10^{-6}$ ph cm$^{-2}$ s$^{-1}$, with EW against the continuum in eV. The improvement in the 1-5 keV fit is given by $\Delta$ $chi^{2}$ for the 
addition of each absorption line
in turn}
\begin{tabular}{@{}lcccccc@{}}
\hline
Line & Energy$_{obs}$ & Energy$_{lab}$ &  ratio & line flux & EW  &  $\Delta chi^{2}$\\
\hline
NeX Ly$\alpha$ & 1.078$\pm$0.05 & 1.022 & 1.055 & -3.9$\pm$2.2 & 5 & 7  \\
MgXI 1s-2p & 1.434$\pm$0.14 & 1.352 & 1.061 & -3.4$\pm$2 & 7 & 5 \\
MgXII Ly$\alpha$ & 1.546$\pm$0.02 & 1.473 & 1.050 & -6.4$\pm$1.5 & 16 & 29 \\
SiXIII 1s-2p & 2.003$\pm$0.02 & 1.866 & 1.073 & -4.5$\pm$1.3 & 17 & 12 \\
SXV 1s-2p & 2.625$\pm$0.06 & 2.460 & 1.067 & -3.3$\pm$1.9 & 21 & 10 \\
SXVI Ly$\alpha$ & 2.765$\pm$0.03 & 2.620 & 1.055 & -3.8$\pm$1.1 & 26 & 20 \\
\hline
\end{tabular}
\end{table*}

With the higher energy resolution of the MOS camera the absorption features at $\sim$1.47 keV and
$\sim$2.71 keV now both appear as a resolved line pair, with the energy spacing of the respective He$\alpha$ and Ly$\alpha$ resonance lines of Mg and S.
Furthermore, additional narrow absorption features are seen to match with same K-shell resonance lines of Ne, Si and possibly Ar.

To quantify these absorption features we explored the MOS data with Xspec. We first fitted the MOS data at 1-5 keV with a power law to provide a baseline; below 1 keV the spectrum rises
steeply due to strong soft X-ray emission (Pounds and Reeves 2006).  Several narrow
features clearly visible in the data-to-power-law-model ratio plot (figure 3, upper panel) contributed to a poor statistical fit ($\chi^{2}$=351/266). Fitting gaussians
to the visible features, with a fixed width of $\sigma$= 10eV, found 6 significant negative (absorption) lines. The overall improvement to the fit was very significant with 
$\chi^{2}$ reduced to 270/254. The fitted line energies and fluxes are listed in Table 1
where the observed line energy is in each case compared with the most likely identification, chosen as the nearest resonance transition of an abundant ion. Crucially, 
all 6
line energies exhibit a `blue shift' in the range $\sim$5-7 \%. Assuming the same ratio for the absorption line observed at $\sim$7.07 keV gives a preferred identification
with He$\alpha$ of FeXXV (figure 2).
In the rest frame of \pg\ the revised identification of the absorption spectrum now yields an increased outflow velocity in the range v$\sim$0.13-0.15c.  

\section{Comparison with an ionised outflow}

To test the compatibility of the visual absorption line set with a physical absorber we then compared the MOS data with a photoionised gas modelled 
using the XSTAR code (Kallman et al 1996). Free parameters of the absorber in this comparison were the column density and ionisation 
parameter, with
outflow (or inflow) velocities included as an adjustment to the apparent redshift of the absorbing gas. All relevant abundant elements from Ne to Fe
were included with the relative abundances constrained to within a factor 2 of solar. 

Since our primary aim was to check the energies and relative strength of the principal absorption lines identified in the visual spectral fit
shown in figure 2, the model did not attempt to match the broad excess flux near $\sim$6 keV; it therefore consisted only of a power law with
photoionised absorber. Fitting over the 1-10 keV band the addition of the photoionised absorber improved the spectral fit from
$\chi^{2}$ of 522 for 358 degrees of freedom to 465/345. The best-fit column density was N$_{H}$$\sim$$2\times 10^{22}$ cm$^{-2}$, with an ionisation parameter of
log$\xi$=2.9$\pm$0.4 and nominal relative abundances of Ne, Mg, Si, S, Ar and Fe of 0.5, 1, 0.5, 1, 0.5 and 0.5. Figure 3 (mid panel) reproduces this absorbed power law model, with the
strongest predicted absorption lines (in order of increasing energy) corresponding to K-shell resonance transitions of Ne, Mg, Si, S, Ar and Fe,
supporting the visual assessment of figure 2. Additionally, we note the Ly$\beta$ line of MgXII would occur at
$\sim$1.83 keV, suggesting much of the observed deficit at that energy may be real (and that our MOS calibration models the neutral Si edge rather well).

\begin{figure}                                                          
\centering
\includegraphics[width=4.3cm, angle=270]{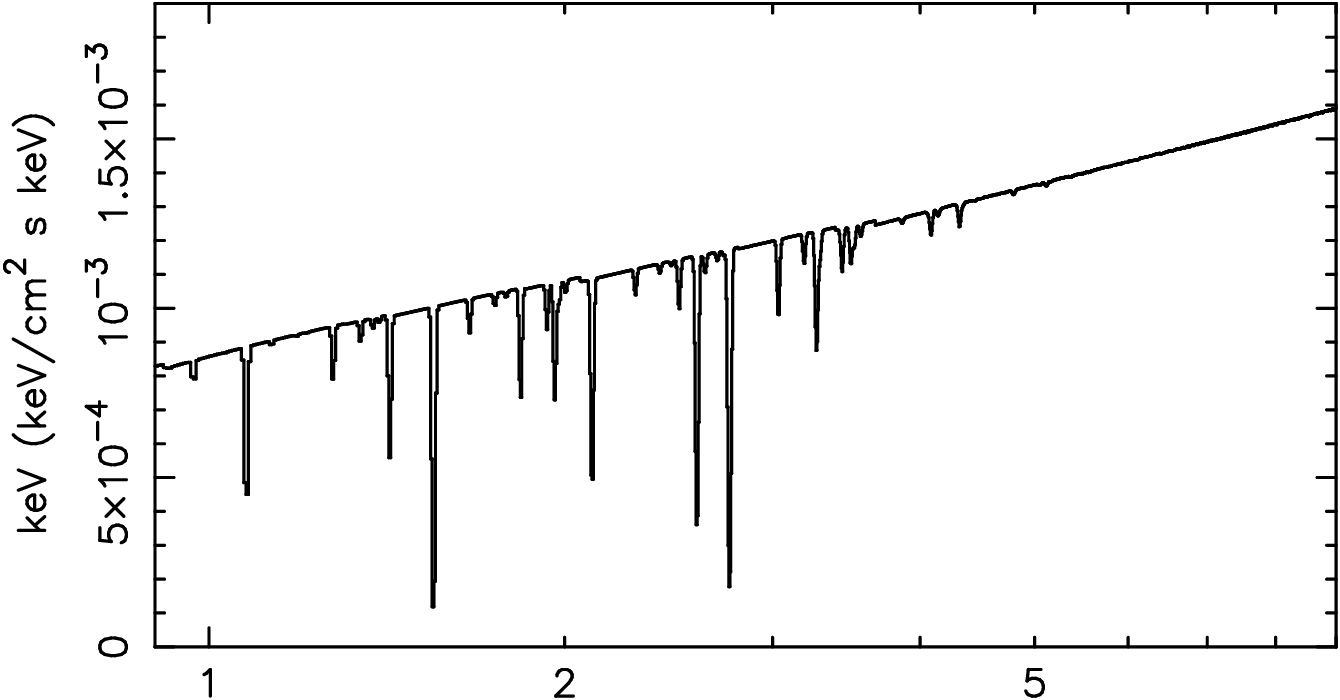}                                          
\centering                                                              
\includegraphics[width=4.3cm, angle=270]{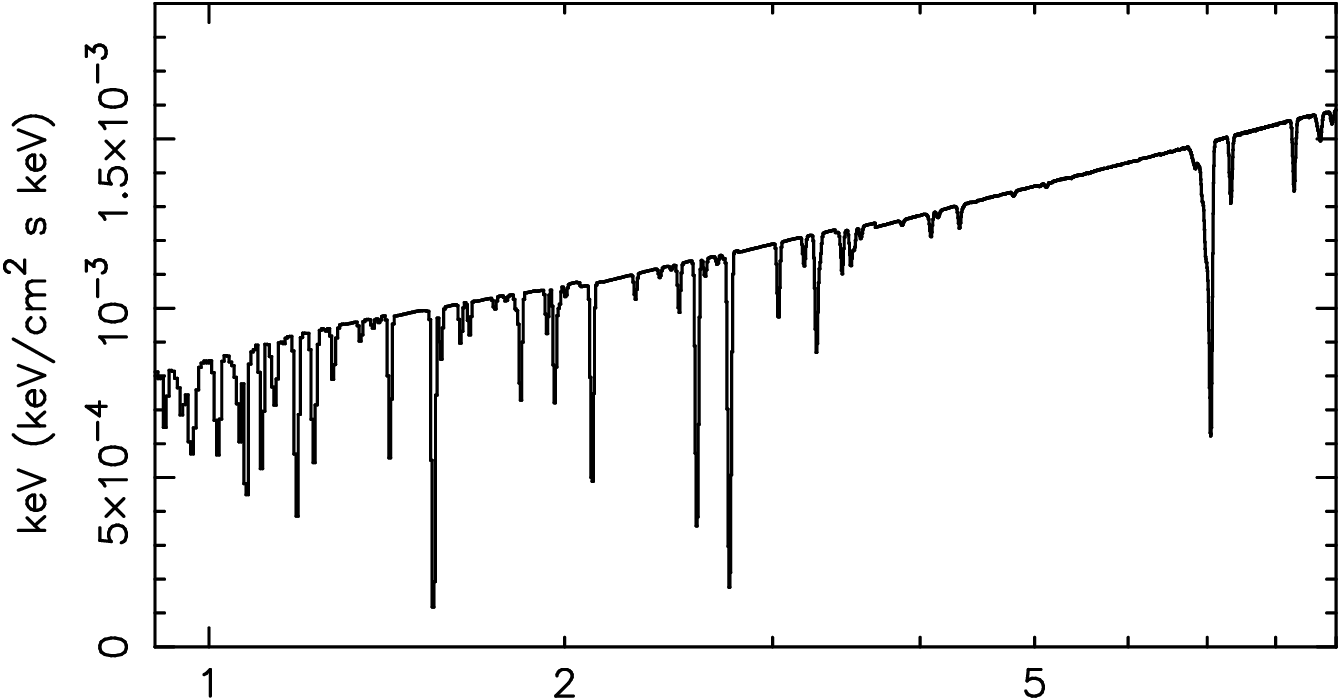}                                          
\centering                                                              
\includegraphics[width=4.7cm, angle=270]{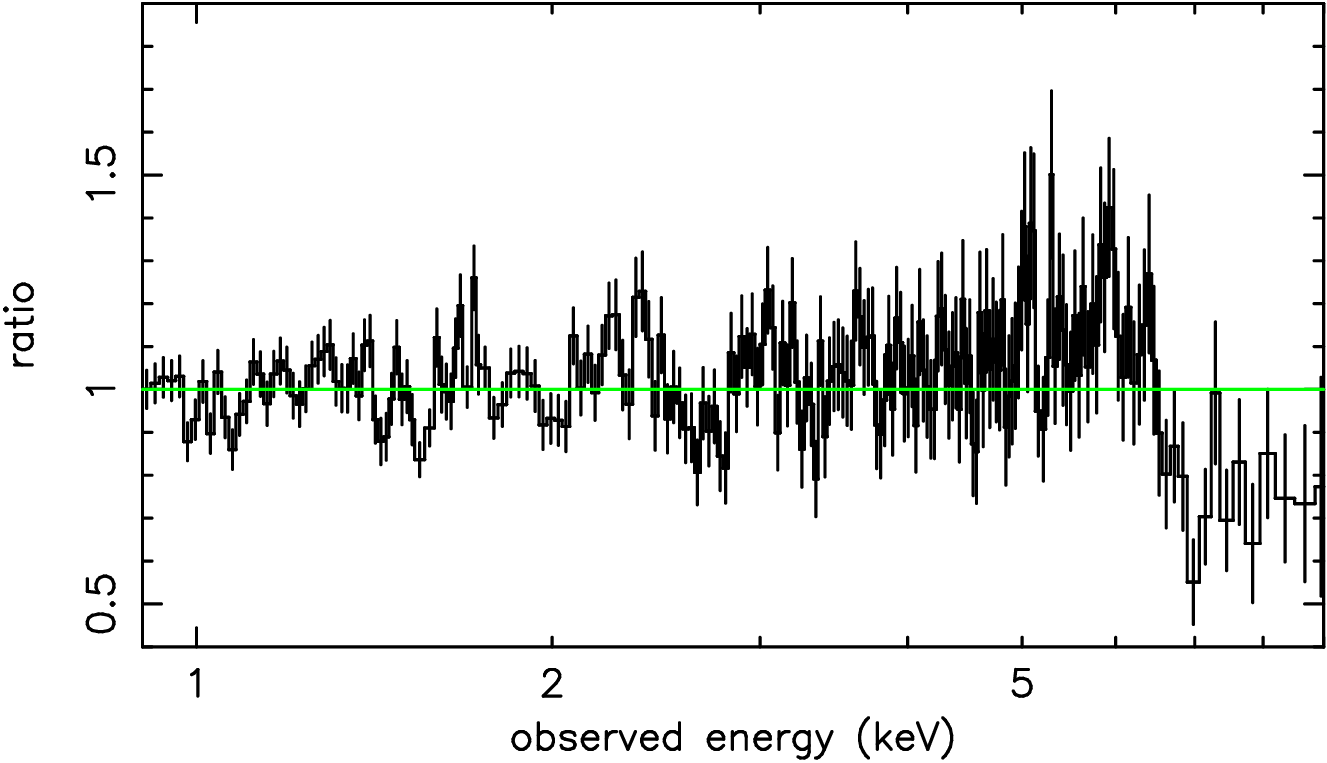}                                          
\caption                                                                
{(top) Photoionised absorber model as described in the text, with principal absorption lines, in order of increasing energy, of Ne, Mg, Si, S
and Ar, but with no Fe;                                                               
(middle) photoionised absorber as above but with Fe now included; (bottom) repeat of the data-to-power law model ratio shown in figure 2;}      
\end{figure}

To clarify the comparison with the MOS data in the lower energy band, the upper panel of figure 3 illustrates the photoionised absorber with the Fe abundance 
set to zero. Removing Fe
from the absorber also shows that potential confusion from
Fe-L absorption is limited above $\sim$1.2 keV. The apparent redshift read from the XSTAR model was (-4.9$\pm$0.3) x 10$^{-2}$, 
corresponding to an outflow velocity (in the rest frame of \pg) of v$\sim$0.130$\pm$0.003c. 
While this value is consistent with the range of velocities of 0.13-0.15c deduced from the individual line fitting in Xspec, the implied precision is probably not justified
given the simplicity of the single absorber XSTAR model.

Although the addition of a lower ionisation absorber would improve the match to the observed line ratios, in particular in providing stronger absorption for He-like
Mg and Ne, we have not included a second absorber in the model, having already
achieved our objective of modelling a high velocity outflow across multiple ions. However, it is interesting to note that the addition of a lower ionisation
component
would also enhance the `red wing' to the Fe K absorption, already showing up in the XSTAR plot in figure 3 (mid panel), due to increased contributions from ions of FeXX-XXIV, 
and thereby explain the
apparently resolved $\sim$7 keV absorption line ($\sigma$=168$\pm$46 eV) in the pn spectrum. 

In summary, modelling the MOS absorption spectrum with a photoionised absorber shows it to be physically compatible with a highly ionised outflow of velocity in the range
0.13-0.15c, as derived from individual line fitting.

\subsection{A revised mass rate and energy in the high velocity flow}

In their previous analysis of the \pg\ absorption line spectrum P03 showed that, provided the high velocity outflow was not tightly collimated, then the mass
loss rate and mechanical energy in the flow would be comparable to the mass accretion rate and $\sim$10 \% of the bolometric luminosity. We can now
re-estimate those quantities. In doing so we use an estimate of 7\% for the covering factor of the fast outflow, calculated by comparing the absorbed and
re-emitted power in the highly ionised gas derived from a broad band fit to the full 0.3-10.0 keV spectrum of \pg\ (Pounds and Reeves 2006).

Assuming a radial flow, the outflow mass rate is then $\mo$ = 0.3$\pi$.nr$^{2}$.v.m$_{p}$ $\sim$0.2$\pi$.L$_{ion}$/$\xi$.v.m$_{p}$ where n is the particle density at a
radius r, and L$_{ion}$$\sim~10^{44}$~erg s$^{-1}$ is the ionising luminosity. With the observed ionisation parameter log$\xi$$\sim$3, we find $\mo$
$\sim$$2.5\times 10^{26}$ gm s$^{-1}$ ($\sim$3.5 $\msun$ yr$^{-1}$).

The corresponding mechanical energy in the fast, highly ionised outflow is then $\sim$$2\times 10^{45}$ erg s$^{-1}$, compared with an estimated bolometric luminosity
for \pg\ of $\sim$$5\times$$ 10^{45}$~erg s$^{-1}$. We note this ratio is a factor $\sim$3 higher than the ratio v/c expected from a radiation driven
wind (King and Pounds 2003). However, given the undoubted simplification of our single-absorber model, and uncertainties in derivation of the covering
factor, it is probably premature to adopt the popular appeal to magnetic forces to drive the outflow! 

\section{How common are energetic, high velocity flows?}

The Black Hole Winds model of King and Pounds (2003) provided a simple physical basis whereby massive, high velocity outflows can be
expected in AGN accreting near the Eddington limit. A simultaneous observation of \pg\ in 2001 with the
\xmm\ Optical Monitor (Mason \et\ 2001) showed the energetically  dominant BBB to be at a typical value, indicating a bolometric luminosity of
$\sim$$5\times$$ 10^{45}$~erg s$^{-1}$. With a reverberation mass estimate for the SMBH in \pg\ of $M \sim 4 \times 10^{7}\Msun$ (Kaspi \et\ 2000)
that luminosity suggests accretion in \pg\ is indeed near the Eddington rate, a conclusion consistent with the optical description of \pg\
as a Narrow Line QSO (Boroson and Green 1992, Kaspi \et\ 2000), given that a high accretion ratio has been causally linked with Narrow Line
Seyfert 1 galaxies (eg Pounds and Vaughan 2000 and references therein). It may well be that energetic outflows are a signature of
Eddington-limited accretion in AGN. This might be the case for the bright Seyfert 1 IC4329A, which Markowitz \et (2006) have recently
shown to exhibit a strongly blue-shifted Fe K absorption line indicating a highly ionised outflow at v$\sim$0.1c. While relatively rare in the local 
universe such X-ray spectra could be common for
luminous, higher redshift AGN.

\section{Summary}

(1) A previous analysis of the 2001 \xmm\ observation of the bright quasar \pg\ reported evidence of a high velocity ionised outflow, with a mass
and kinetic energy comparable to the accretion mass and bolometric luminosity, respectively (P03).

(2) This finding is now confirmed, with the previous uncertainty in the derived velocity removed by securing the identification of the main
observed absorption lines.

(4) We suggest that fast, energetic outflows may be a typical signature of type 1 AGN accreting at or close to the Eddington  limit.

\section*{ Acknowledgements }
The results reported here are based on observations obtained with \xmm, an ESA science mission with
instruments and contributions directly funded by ESA Member States and
the USA (NASA).
The authors wish to thank the SOC and SSC teams for organising the \xmm\
observations and initial data reduction. We also thank the anonymous referee for a careful reading of the text and constructive comments.

\end{document}